\documentclass{PoS}

\title{Effective gluon potential and Yang-Mills thermodynamics}

\ShortTitle{Effective gluon potential and YM thermodynamics}

\author{\speaker{Chihiro Sasaki}\\
        Frankfurt Institute for Advanced Studies,
        D-60438 Frankfurt am Main,
        Germany\\
        E-mail: \email{sasaki@fias.uni-frankfurt.de}}

\author{Krzysztof Redlich\\
        Institute of Theoretical Physics, University of Wroclaw,
        PL-50204 Wroc\l aw,
        Poland\\
        E-mail: \email{krzysztof.redlich@ift.uni.wroc.pl}}

\abstract{ 
We derive the Polyakov-loop thermodynamic potential  in the 
perturbative approach to pure SU(3) Yang-Mills theory.  
The potential expressed in terms of the Polyakov loop in the fundamental
representation corresponds to that of the strong-coupling expansion,
of which the relevant coefficients of the gluon energy distribution are
specified  by characters of the SU(3) group.  
At high temperature,  the potential
exhibits the correct asymptotic behavior, whereas at low
temperature, it disfavors gluons as appropriate dynamical
degrees of freedom. To quantify the Yang-Mills thermodynamics
in confined phase, we introduce  a hybrid approach which matches
the effective gluon potential to that  of glueballs,  
constrained by the QCD trace anomaly in terms of dilaton fields.
}

\FullConference{Xth Quark Confinement and the Hadron Spectrum,\\
		October 8-12, 2012\\
		TUM Campus Garching, Munich, Germany}

\begin{document}

\section{Introduction}

The structure of the QCD phase diagram and thermodynamics at
finite baryon density is of crucial importance in heavy-ion
phenomenology. Due to the sign problem in lattice calculations,
a major approach to a finite density QCD is based on effective
Lagrangians possessing the same global symmetries as the underlying QCD.
The $SU(N_c)$ Yang-Mills theory has a global $Z(N_c)$ symmetry
which is dynamically broken at high temperature. This is characterized by
the Polyakov loop that plays a role of an order parameter of the $Z(N_c)$
symmetry~\cite{mclerran}. Effective models for the Polyakov loop were
suggested as a macroscopic approach to the pure gauge
theory~\cite{pisarski,pisarski:lect}. 
Their  thermodynamics is qualitatively
in agreement with that obtained in lattice gauge theories~\cite{lat:eos}.
Alternative approaches are based on the quasi-particle picture of
thermal gluons~\cite{peshier}. When gluons propagating in
a constant gluon background are considered, the quasi-particle models 
naturally merge with the Polyakov loops, that appear in the partition 
function, as characters of 
the color gauge group~\cite{turko,meisinger:2002,meisinger,kusaka,slqcd}.

In this contribution we show, that  the $SU(3)$ gluon thermodynamic
potential can be derived directly  from the Yang-Mills theory and 
is expressed  in terms of the Polyakov loops in the fundamental 
representation. 
We  summarize  its properties and argue that at hight temperatures,  
it  exhibits the correct asymptotic behavior,
whereas at low temperatures,  it disfavors gluons~\cite{hybrid}.
We therefore suggest a hybrid approach to  Yang-Mills  thermodynamics,
which combines the  effective gluon potential with
glueballs implemented as dilaton fields.

\section{Thermodynamics of hot gluons}
We start from the partition function of the pure Yang-Mills theory
\begin{equation}
Z =
\int{\mathcal D}A_\mu{\mathcal D}C{\mathcal D}\bar{C}\,
\exp\left[i\int d^4x {\mathcal L}_{\rm YM}\right]\,,
\end{equation}
with gluon $A_\mu$ and ghost $C$ fields.
Following~\cite{pisarski:lect,GPY} we employ the background field method
to evaluate  the functional integral. The gluon field is decomposed into
the background $\bar{A}_\mu$ and the quantum  $\check{A}_\mu$ fields,
\begin{equation}
A_\mu = \bar{A}_\mu + g\check{A}_\mu\,.
\end{equation}
The partition function is arranged as
\begin{eqnarray}
\ln Z
= V\int\frac{d^3 p}{(2\pi)^3}
\ln\det\left( 1 - \hat{L}_A e^{-|\vec{p}|/T}\right)
{}+ \ln M(\phi_1,\phi_2)\,,
\label{parti}
\end{eqnarray}
where $\hat{L}_A$ is the Polyakov loop matrix in the adjoint 
representation and 
the two  angular variables,  $\phi_1$ and $\phi_2$,  
represent the rank of the $SU(3)$ group.
The $M(\phi_1,\phi_2)$ is the Haar measure 
\begin{eqnarray}
\label{haar}
M(\phi_1,\phi_2)
&=&
\frac{8}{9\pi^2}\sin^2\left( \frac{\phi_1 - \phi_2}{2}\right)
\sin^2\left( \frac{2\phi_1 + \phi_2}{2}\right)
\sin^2\left( \frac{\phi_1 + 2\phi_2}{2}\right)\,,
\end{eqnarray}
for a fixed volume $V$,  which is normalized such that
\begin{equation}
\int_0^{2\pi}\int_0^{2\pi}d\phi_1 d\phi_2 M(\phi_1,\phi_2) = 1\,.
\end{equation}
The first term in Eq.~(\ref{parti}) yields the gluon thermodynamic
potential
\begin{equation}
\Omega_g
= 2T \int\frac{d^3p}{(2\pi)^3}\mbox{tr}\ln
\left( 1 - \hat{L}_A\, e^{-E_g/T} \right)\,,
\label{omega0}
\end{equation}
where  $E_g = \sqrt{|\vec{p}|^2 + M_g^2}$ is the quasi-gluon energy 
and the effective gluon mass $M_g$ is introduced from  phenomenological
reasons.

We define the gauge invariant quantities from the Polyakov loop
matrix in the fundamental representation $\hat{L}_F$,  as
\begin{eqnarray}
\Phi = \frac{1}{3}\mbox{tr}\hat{L}_F\,,
\quad
\bar{\Phi} = \frac{1}{3}\mbox{tr}\hat{L}_F^\dagger\,.
\end{eqnarray}
Performing the trace over colors and expressing it
in terms of $\Phi$ and its conjugate $\bar\Phi$,
one arrives at
\begin{equation}
\Omega_g
= 2T \int\frac{d^3p}{(2\pi)^3}\ln
\left( 1 + \sum_{n=1}^8C_n\, e^{-nE_g/T}
\right)\,,
\label{gluon}
\end{equation}
with the coefficients $C_n$, 
\begin{eqnarray}
\label{coeff}
C_8
&=&
1\,,
\nonumber\\
C_1
&=&
C_7
= 1 - 9\bar{\Phi}\Phi\,,
\nonumber\\
C_2
&=&
C_6
= 1 - 27\bar{\Phi}\Phi
{}+ 27\left( \bar{\Phi}^3 + \Phi^3\right)\,,
\nonumber\\
C_3
&=&
C_5
= -2 + 27\bar{\Phi}\Phi
{}- 81\left( \bar{\Phi}\Phi \right)^2\,,
\nonumber\\
C_4
&=&
2\left[
-1 + 9\bar{\Phi}\Phi - 27\left( \bar{\Phi}^3 + \Phi^3\right)
{}+ 81\left( \bar{\Phi}\Phi \right)^2
\right]\,.
\end{eqnarray}
Thus, the gluon energy distribution is identified solely by the
characters of the fundamental and the conjugate representations of
the $SU(3)$ gauge group.

We introduce an effective thermodynamic potential in the large volume
limit from Eq.~(\ref{parti}) as follows:
\begin{eqnarray}
\label{full}
\Omega
&=&
\Omega_g + \Omega_\Phi + c_0\,,
\end{eqnarray}
where $\Omega_g $ is given by Eq.~(\ref{gluon}) and
the Haar measure part is found as
\begin{eqnarray}
\Omega_\Phi
&=&
-a_0T\ln\left[ 1 - 6\bar{\Phi}\Phi + 4\left( \Phi^3 + \bar{\Phi}^3\right)
{}- 3\left(\bar{\Phi}\Phi\right)^2\right]\,.
\label{dec}
\end{eqnarray}
The potential (\ref{full}) has, in general,  three free parameters;  
$a_0$,  $c_0$ and the gluon mass $M_g$.
They can be chosen e.g. to reproduce the equation of state
obtained in lattice gauge theories.
It is straightforward to see,  that the result of a non-interacting boson
gas is recovered at asymptotically high temperature.
Indeed, taking $\Phi, \bar{\Phi} \to 1$ one finds
\begin{equation}
\label{expansion}
\Omega_g(\Phi=\bar{\Phi}=1)
= 16T \int\frac{d^3p}{(2\pi)^3}
\ln\left( 1 - e^{-E_g/T} \right)\,.
\end{equation}
On the other hand, for a sufficiently large $M_g/T$, as expected near
the phase transition, one can approximate the potential as
\begin{equation}
\Omega_g
\simeq {\frac{T^2M_g^2}{\pi^2}}\sum_{n=1}^8{\frac{C_n}{n}}K_2(n\beta M_g)\,,
\label{effective}
\end{equation}
with the Bessel function $K_2(x)$.
In the quasi-particle approach, the above  result  can also be considered 
as a strong-coupling expansion,  regarding the
relation $M_g(T)=g(T)T$ with an effective gauge coupling $g(T)$.

The effective action to the  next-to-leading order of the strong coupling
expansion  is  obtained in terms of group characters as~\cite{slqcd}, 
\begin{equation}
S_{\rm eff}^{\rm (SC)}
= \lambda_{10}S_{10} + \lambda_{20}S_{20} + \lambda_{11}S_{11}
{}+ \lambda_{21}S_{21}\,,
\end{equation}
with products of  characters $S_{pq}$, specified by two integers
$p$ and $q$ counting the numbers of fundamental and conjugate
representations, and couplings $\lambda_{pq}$ being real functions of
temperature.
Making the character expansion of Eq.~(\ref{effective}),
one readily finds the correspondence between $S_{pq}$ and $C_n$ as
\begin{eqnarray}
C_{1,7} = S_{10}\,,
\quad
C_{2,6} = S_{21}\,,
\quad
C_{3,5} =S_{11}\,,
\quad
C_4 = S_{20}\,.
\end{eqnarray}

On the other hand, taking the leading contribution,  $\exp[-M_g/T]$ 
in the expansion, the ``minimal model'' is deduced with
\begin{equation}
\Omega_g
\simeq -{\mathcal F}(T,M_g)\bar{\Phi}\Phi\,,
\label{approx}
\end{equation}
where the negative sign is required for a first-order
transition~\cite{slqcd}. The function ${\mathcal F}$ can be extracted from
Eq.~(\ref{full}) and the resulting  potential is of the form 
widely used in the PNJL model~\cite{pnjl,pnjl:log,pnjl:sus,pnjl:poly}.

\section{A hybrid approach}

Although the potential (\ref{full}) describes quite well thermodynamics in 
deconfined phase, it totally fails in the confined phase.
In the confined phase,  $\langle\Phi\rangle=0$ is dynamically favored by
the ground state,   thus the $C_1=1$ term remains as the main 
contribution. Consequently
\begin{equation}
\Omega_g(\Phi=\bar{\Phi}=0)
\simeq 2T \int\frac{d^3p}{(2\pi)^3}
\ln\left( 1 + e^{-E_g/T} \right)\,.
\label{lowT}
\end{equation}
One clearly sees that $\Omega_g$ does not posses the correct sign in
front of $\exp[-E_g/T]$,  expected from the Bose-Einstein statistics.
This implies that  the entropy and the energy densities are {\it negative}.
On the other hand, if one  uses  the approximated  form (\ref{approx}), 
the pressure vanishes  at any temperature below $T_c$.
Obviously,  this is an  unphysical behavior  since there exist 
color-singlet states, i.e. glueballs, contributing to thermodynamics and
they must generate a non-vanishing pressure.

This aspect is in a striking contrast to the quark sector. 
The thermodynamic potential for  quarks  and anti-quarks
with $N_f$ flavors is obtained as~\cite{pnjl,megias}
\begin{eqnarray}
\Omega_{q+\bar q}
=
-2N_f T \int\frac{d^3p}{(2\pi)^3}
\ln
\left[ 1 + N_c\left( \Phi + \bar{\Phi}e^{-E^+/T}\right)e^{-E^+/T}
{}+ e^{-3E^+/T}\right]
{}+
\left(\mu \to -\mu\right) \,,
\label{omega:quark}
\end{eqnarray}
with $E^\pm = E_q \mp \mu$ being  the energy of a quark or anti-quark.
In the limit,  $\Phi,\bar{\Phi}\to 0$,
the one- and two-quark states are suppressed and only the three-quark 
(``baryonic'') states,  $\sim\exp(-3E^{(\pm)}/T)$, survives. This, 
on a qualitative level, is similar to confinement properties in QCD 
thermodynamics~\cite{pnjl:sus}.
One should, however,  keep in mind, that
such quark models  yield only  colored quarks being  {\it statistically}
suppressed at low temperatures.
On the other hand, unphysical thermodynamics below $T_c$ described by
the gluon sector (\ref{full}) apparently indicates,  that  gluons
are {\it physically} forbidden.
Interestingly, this property is not spoiled by the presence of quarks.
Indeed, in this case and  at  $T < T_c$  the thermodynamic potential is 
approximated as
\begin{equation}
\Omega_g + \Omega_{q+\bar{q}}
\simeq
\frac{T^2}{\pi^2}\left[
M_g^2 K_2\left(\frac{M_g}{T}\right)
{}- \frac{2N_f}{3}M_q^2 K_2\left(\frac{3M_q}{T}\right)
\right]\,.
\label{lowTapprox}
\end{equation}
Assuming that  glueballs and nucleons are made from two weakly-interacting
massive gluons and three massive quarks respectively and  putting empirical
numbers,  $M_{\rm glueball}=1.7$ GeV and $M_{\rm nucleon}=0.94$ GeV,
one finds that   $M_g = 0.85$ GeV and $M_q = 0.31$ GeV.
Substituting these mass values in Eq.~(\ref{lowTapprox}),  one still gets 
the negative entropy density 
at any temperature and   for either $N_f=2$ or $N_f=3$,  as found 
in the pure Yang-Mills theory.

The  unphysical equation of state (EoS) in confined phase can be avoided,
when gluon degrees of freedom are replaced with glueballs.
A glueball is introduced as a dilaton field $\chi$ representing the gluon
composite $\langle A_{\mu\nu}A^{\mu\nu}\rangle$,  which is responsible
for the QCD trace anomaly~\cite{schechter}. The Lagrangian is of
the standard form,
\begin{eqnarray}
\label{dilaton}
{\mathcal L}_\chi
=
\frac{1}{2}\partial_\mu\chi\partial^\mu\chi - V_\chi\,,
\quad
V_\chi
=
\frac{B}{4}\left(\frac{\chi}{\chi_0}\right)^4
\left[ \ln\left(\frac{\chi}{\chi_0}\right)^4 - 1 \right]\,,
\end{eqnarray}
with the bag constant $B$ and a dimensionful quantity $\chi_0$,
to be fixed from the vacuum energy density and the glueball mass.
One readily finds the thermodynamic potential of the glueballs as
\begin{eqnarray}
\Omega
&=&
\Omega_\chi + V_\chi + \frac{B}{4}\,,
\quad
\Omega_\chi
=
T\int\frac{d^3p}{(2\pi)^3}\ln\left(1-e^{-E_\chi/T}\right)\,,
\nonumber\\
E_\chi
&=&
\sqrt{|\vec{p}|^2 + M_\chi^2}\,,
\quad
M_\chi^2 = \frac{\partial^2 V_\chi}{\partial\chi^2}\,,
\label{conf}
\end{eqnarray}
where a constant $B/4$ is added so that $\Omega = 0$ at zero temperature.

We propose the following hybrid approach which accounts for gluons
and glueballs degrees of freedom by combining Eqs.~(\ref{full}) and
(\ref{conf}),
\begin{equation}
\Omega = \Theta(T_c-T)\,\Omega(\chi) + \Theta(T-T_c)\,\Omega(\Phi)\,.
\label{model}
\end{equation}
For a given $M_g$,  the model parameters, $a_0$ and $c_0$, are fixed
by requiring,  that
 $\Omega(\Phi)$ yields a first-order phase transition at $T_c=270$ MeV
and that 
 $\Omega(\chi)$ and $\Omega(\Phi)$ match at $T_c$.
The resulting EoS follows general trends seen in lattice data~\cite{hybrid}.
The model can be improved further by introducing a thermal gluon mass, 
$M_g(T) \sim g(T)T$, as carried out e.g. in \cite{meisinger}.

\section{Summary}

We have derived the thermodynamic potential in the $SU(3)$  Yang-Mills
theory in the presence of a uniform gluon background field. 
The potential accounts for quantum statistics and reproduces 
an  ideal gas limit at high temperature.
Within the character expansion, the  one-to-one correspondence to the 
effective action in the strong-coupling expansion is obtained.
Different effective potentials used so far appear as limiting cases of
our result.

The phenomenological consequence is that
gluons are disfavored as appropriate degrees of freedom in confined phase. 
This property is in remarkable contrast to the
description of ``confinement'' within a class of chiral models with
Polyakov loops~\cite{pnjl,pnjl:poly}, where colored quarks are activated
at any temperature.
Further investigations of the $SU(3)$ gluodynamics guided by available
lattice results with the effective gluon mass and
a more realistic description of an effective  QCD thermodynamics
with quarks are desired.

\section*{Acknowledgments}

C.~S. acknowledges partial support by the Hessian
LOEWE initiative through the Helmholtz International
Center for FAIR (HIC for FAIR).
K.R. acknowledges support by the Polish Science Foundation (NCN).


\end{document}